\documentclass[a4paper, 10pt,twocolumn]{article}

\usepackage{amssymb}
\usepackage{amsmath}
\usepackage{graphicx}

\begin{document}

\title{Low energy Langmuir cavitons:\\ 
asymptotic limit of weak turbulence}

\author{P.~Henri$^{1,2}$, F.~Califano$^{1,2}$, C.~Briand$^{2}$, A.~Mangeney$^{2}$ \\
\tiny{$^{1}$ Dip. Fisica, Universit\`{a} di Pisa; Largo Pontecorvo 3, 56127 Pisa, Italy}  \\
\tiny{$^{2}$ LESIA, Observatoire de Paris, CNRS, UPMC, Universit\'e Paris Diderot; 5 Place J.~Janssen, 92190 Meudon, France}  }

\date{}

\twocolumn[
  \begin{@twocolumnfalse}
    \maketitle
    \begin{abstract}
Langmuir turbulence is an archetype of wave turbulence in plasma physics. By means of 1D-1V Vlasov-Poisson simulations, we show that coherent structures, called Langmuir cavitons, are generated by the long time evolution of Langmuir weak turbulence, thus illustrating the breakdown of a weak turbulence regime. 
These structures correspond to an equilibrium between the pressure forces and the ponderomotive force resulting from high frequency Langmuir oscillations. Langmuir cavitons are typical features of strong Langmuir turbulence expected to be generated at high energy and to saturate when Langmuir energy is of the order of the plasma thermal energy. 
Despite this wide-spread belief, here we observe that cavitons, emerging from weak Langmuir turbulence evolution, saturate at much lower energies. We show that these Langmuir coherent structures are characterized by a much larger length scale with respect to the Debye length. This gives evidence that "large" and "shallow" stable cavitons should be seen in space plasma observations. The transition toward strong turbulence is shown to be a consequence of an initial weak turbulent inverse cascade. 
Finally, the effective equation of state for ion acoustic oscillations is tested numerically from the kinetic model. \\
    \end{abstract}
  \end{@twocolumnfalse}
]

 \section{Introduction}
The nonlinear evolution of waves is usually classified in terms of "weak" and "strong" turbulence. 
The meaning of these terms is not well established although the major conceptual difference between weak and strong turbulence is the presence of a characteristic dimensionless parameter $\varepsilon$ characterizing the level of nonlinearity, as the ratio between a typical linear (or dispersive) time (the electron plasma period in Langmuir turbulence) and the nonlinear time, depending on the wave amplitudes. Turbulence is  then considered to be "weak" if $\varepsilon \ll 1$; otherwise it is "strong". 
A second assumption is usually made: weak turbulence results from a superposition of finite, but weak amplitude waves, obeying the linear dispersion relation, but with randomly distributed phases. Starting with (almost) random phases, such randomness must be preserved over the nonlinear evolution time. 
Two key points are decisive for the validity of Random Phase Approximation (RPA): 
(a) the amplitude of the fields and (b) the bandwidth of the phenomena under consideration \cite{Sagdeev&Galeev1969,Goldman1984RvMP,Robinson1997RvMP}. 
At finite but small energy, the nonlinear dynamics is described by three-waves or four-waves interactions in the RPA through the kinetic wave equations, whereas at higher energies intermittency dominates the dynamics, through the apparition of coherent structures. 

To summarize, weak turbulence is based on the following points: (i) linear dispersion still holds; (ii) statistical homogeneity in space holds; (iii) wave particle interaction is described by resonant quasilinear theory. 
On the other hand, strong turbulence is based on the following: (i) linear dispersion no longer holds; (ii) strong statistical inhomogeneity in space; (iii) wave particle interaction become complex because of  wave (in density cavities) and particle trapping. \\

Only a few attempts have been made to compare weak turbulence theory with numerical results. The three dimensional equations for capillary water waves has been solved numerically \cite{Pushkarev&Zakharov1996PRL} and a Zakharov $\&$ Filonenko power-law spectrum \cite{Zakharov&Filonenko1967} was observed. Weak turbulence theory may be valid in some spectral range but fails in others \cite{BivenConnaughtonNewell2003PhyD}. 
Recently, the problem of breakdown of weak turbulence by intermittent events associated with coherent structures has been addressed for different models \cite{BivenNazarenkoNewell2001PhLA, BivenConnaughtonNewell2003PhyD} and an illustration of the modulational instability influence on the breakdown of weak turbulence resulting from an inverse cascade has been given for the nonlinear Schr�dinger equation \cite{Dyachenkoetal1992PhyD}.
However, these results have been established for fluid models, i.e. without kinetic effects, whose role in the phase randomizing is unknown. 
Therefore, the validity of weak turbulence theory in a full kinetic regime is still an open problem to be investigated by numerical simulations. \\

\subsection{Langmuir turbulence}
We hereafter concentrate on the specific case of electrostatic Langmuir turbulence, an archetype of wave turbulence in plasma physics. 
In the weak turbulence regime, the 3-wave (resp. 4-wave) evolution dominates for large (resp. small) wave numbers $k_{L} > k_{MI}$ (resp. $k_{L} < k_{MI}$) through the decay (resp. modulational) instability, which typically transfers the L-wave energy towards smaller (resp. larger) wave vectors. The transition wavenumber is $k_{MI} \lambda_{_{D}} = 1/3 \ c_{s} / v_{th,e}$ with $\lambda_{_{D}}$ the Debye length, $c_{s}$ the ion sound speed and $v_{th,e}$ the electron thermal velocity \cite{Zakharov&al1985PhR}. 
Weak Langmuir turbulence is mainly driven by the electrostatic decay of a Langmuir wave (hereafter L-wave) into another L-wave and an ion acoustic wave (hereafter IA), while several processes associated to the ponderomotive force, as modulational instability, oscillating two stream instability, soliton formation, Langmuir collapse, are considered to dominate the strong turbulence dynamics \cite{Sagdeev&Galeev1969,Goldman1984RvMP}.  
In the specific case of Langmuir turbulence, the turbulence is usually considered "strong" for an electric-to-thermal energy ratio larger than the electron-to-ion mass ratio: $W = \epsilon_{0} E^2 / nk_{B}T >  m_{e}/m_{i}$.

Langmuir cavitons are localized electric fields oscillating at the plasma frequency self-consistently 
associated to density cavities \cite{Zakharov1972JETP} with a large range of variations, $0.01 < \delta n / n < 0.80 $ \cite{Wong1977JPhys}. 
These structures have been observed in the laboratory in several plasma experiments \cite{KimStenzelWong1974PRL} while in natural space plasmas only in active ionospheric experiments \cite{Wong&al1987PRL}. 
Cavitons at "high" energies have also been intensively studied through numerical experiments of the Zakharov equations   \cite{DuboisRoseRussell1991PRL,Guio&Forme2006PhPl,Eliasson&Thide2008JGRA,DoolenDuboisRose1985PRL} and the Vlasov-Poisson equations \cite{Wang&al1996PhPl,Wang&al1997JASTP,SircombeArberDendy2005PhPl}. For moderate forcing, it has been shown that weak turbulence and strong turbulence features can coexist \cite{DuboisRoseRussell1991PRL}. 

The Langmuir cavitons canonical scenario, starting from an initial long wavelength Langmuir spectrum, can be summarized as follows: (a) parametric cascading leading to Langmuir condensation and increase in wave intensity; (b) cavity formation ($\epsilon_{0} E^2/ 2 n k_B T > k^2 \lambda_D^2$); (c) caviton collapse ($k \to \infty$); (d)  particles acceleration and cavity emission of ion sound waves; (e) start of a "caviton nucleation cycle". \\

By considering here the limit of 1D Langmuir turbulence, an archetype of wave turbulence in plasma physics, we show the transition from weak to strong turbulence through the formation of coherent structures (cavitons), independently of the initial level of coherence of Langmuir oscillations. 
The resulting cavitons may saturate at "low" energy levels (electric energy orders of magnitude lower than thermal energy) generally considered to belong to the weak Langmuir turbulence regime and remain then relatively stable. We obtained a power law governing the relation between the scale length of the structures and their energy to be directly tested on space plasma data. Finally, the nonlinear time scale needed to reach the strong turbulence regime is found to scale as the inverse of the initial Langmuir energy. 

These results are obtained with a kinetic description of 1D electrostatic plasma. To our knowledge, no similar work on the breakdown of weak turbulence has been done using a fluid model (Zakharov equations). 

 \section{Model}

We solve the 1D1V Vlasov-Poisson system of equations for the electron and proton distribution function $f_e$, $f_p$ and the self-consistent electric potential and electric field, $\phi$ and $E$. 
All equations are normalized by using electron quantities, the electron charge, mass and thermal velocity, 
$e$, $m_e$ and  $v_{th,e}$, the plasma frequency $\omega_{pe}$, the Debye length $\lambda_{_{D}}$ and a characteristic density and electric field, $\bar n_e$ and $\bar E = m_{e} v_{th,e} \omega_{pe} / e$.

We define $W = 0.5 \times E^2$ as the electric energy density normalized to the electron kinetic energy. Then, the dimensionless Vlasov equations read: 
\begin{equation} \label{eq:vlasovelectrons}
	\frac{\partial f_{e}}{\partial t} + v \frac{\partial f_{e}}{\partial x} - (E+E_{ext}) \frac{\partial f_{e}}{\partial v} = 0 
\end{equation}
\begin{equation} \label{eq:vlasovions}
	\frac{\partial f_{p}}{\partial t} + u \frac{\partial f_{p}}{\partial x} + \frac{1}{\mu} E \frac{\partial f_{p}}{\partial u} = 0
\end{equation}
where $v$ and $u$ are the electron and ion velocity, $\mu = m_e / m_p = 1/1836$ the electron-to-proton mass ratio and $E_{ext}$ an "external" driver acting on the electrons only that can be switched on or off during the runs. 
The details of the external forcing can be found in Appendix 1 of Ref. \cite{Henri&al2010JGR}. 
Finally, the Vlasov equations are self-consistently coupled to the Poisson equation. 
\begin{equation} \label{eq:poisson}
	\frac{\partial^2 \phi}{\partial x^2} = \int f_{e} dv - \int f_{p} du \mathrm{\; \; ; \; \; } E = - \frac{\partial \phi}{\partial x}
\end{equation}
We use a numerical box of length $L_{x} = 5000\, \lambda_{_{D}}$ and a velocity range
$-5 \leq v/v_{th,e} \leq +5$ for electrons and $-5 \leq u/u_{th,i} \leq +5$ protons. 
The spatial and velocity mesh grid is $dx = \lambda_{_{D}}$,  $dv = 0.04\ v_{th,e}$ and $du = 0.04\ u_{th,i}$
where $u_{th,i}$ is the proton thermal velocity. 
Periodic boundary conditions are used in the spatial direction. The initial electron and proton velocity distributions are Maxwellian with equal temperatures, $T_{p} = T_{e}$, a typical condition in solar wind plasma, so that IA fluctuations are efficiently damped out. We 
add at $t=0$ a density random noise in the wavelength range $30 < \lambda < 500$. 

We consider two different initial conditions, (i) coherent and (ii) incoherent L-waves with an initial electric energy level varying in the range $10^{-4} \lesssim W \lesssim 10^{-2}$, corresponding to the transition from weak to strong turbulence regime.
In case (i), the external driver $E_{ext}$, oscillating at $\omega_{pe}$, excites a monochromatic L-wave with wavelength $\lambda_{_{L}}$ and propagating in one direction only. The forcing is switched off when the L-wave reaches the desired amplitude $E_{_{L}}$. 
The initial wave then evolves self consistently according to the Vlasov-Poisson system. 
The corresponding Langmuir energy ranges in the interval $2 \cdot 10^{-4} < W < 10^{-1}$, here again corresponding to the transition from weak to strong Langmuir turbulence regime.
Finally, the initial random density noise is $\delta n / n = 10^{-5}$. 
In case (ii), we exclude the forcing from the beginning and let the system to evolve starting with an electron density random noise corresponding to a flat spectrum in the electric field. The r.m.s. electron density amplitude is $10^{-3} \lesssim \delta n_{e} / n_{e} \lesssim 10^{-2}$, corresponding to a r.m.s. electric field $10^{-2} \lesssim E \lesssim 10^{-1}$, i.e. an energy range $2 \cdot 10^{-4} < W < 2 \cdot 10^{-2}$.

We use a kinetic model to fully take into account all processes, as the wave-particle interactions, that could limit the development of Langmuir turbulence by extracting electric energy and converting it into kinetic energy. The phase velocity $v_{\phi}$ of the initial L-waves is much larger than $v_{th,e}$, so that the Landau damping of Langmuir waves is inefficient. On the other hand, $T_{p} = T_{e}$ (a common situation in space plasmas like the solar wind) so that the generated IA waves are efficiently Landau damped during the transient part of the simulation, corresponding to a weak turbulence evolution. This kinetic damping could limit, first the Langmuir cascade, second, the generation of density fluctuations that are a seed for the subsequent generation of density inhomogeneities observed in the asymptotic part of the simulation, corresponding to the strong turbulence evolution. \\

 \section{Numerical results}

\begin{figure}[t]
\noindent \includegraphics[width=\linewidth]{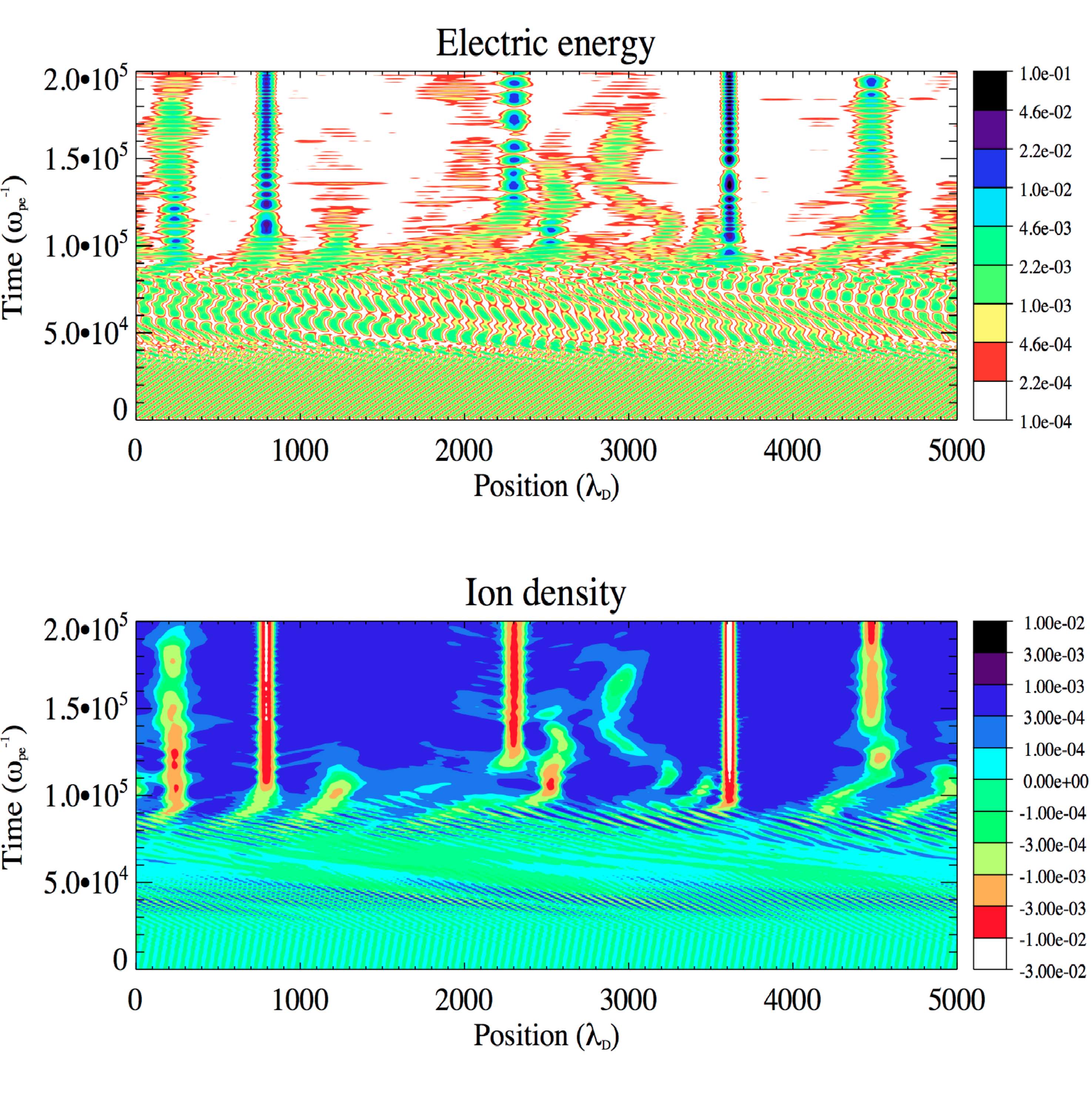}
\caption{Long time evolution of electric energy $W_{_{L}}$ (top panel) and ion density $\delta n / n$ (bottom panel) in the ($x,t$) plane, starting from a monochromatic Langmuir wave. Note in the bottom panel the generation of (i) ion acoustic waves from the Langmuir electrostatic decay between $3\times10^4 < t < 6\times10^4$; (ii) ion cavities filled with electric energy for $t > 10^5$.}
\label{fig1}
\end{figure}

\begin{figure}[t!]
\noindent \includegraphics[width=\linewidth]{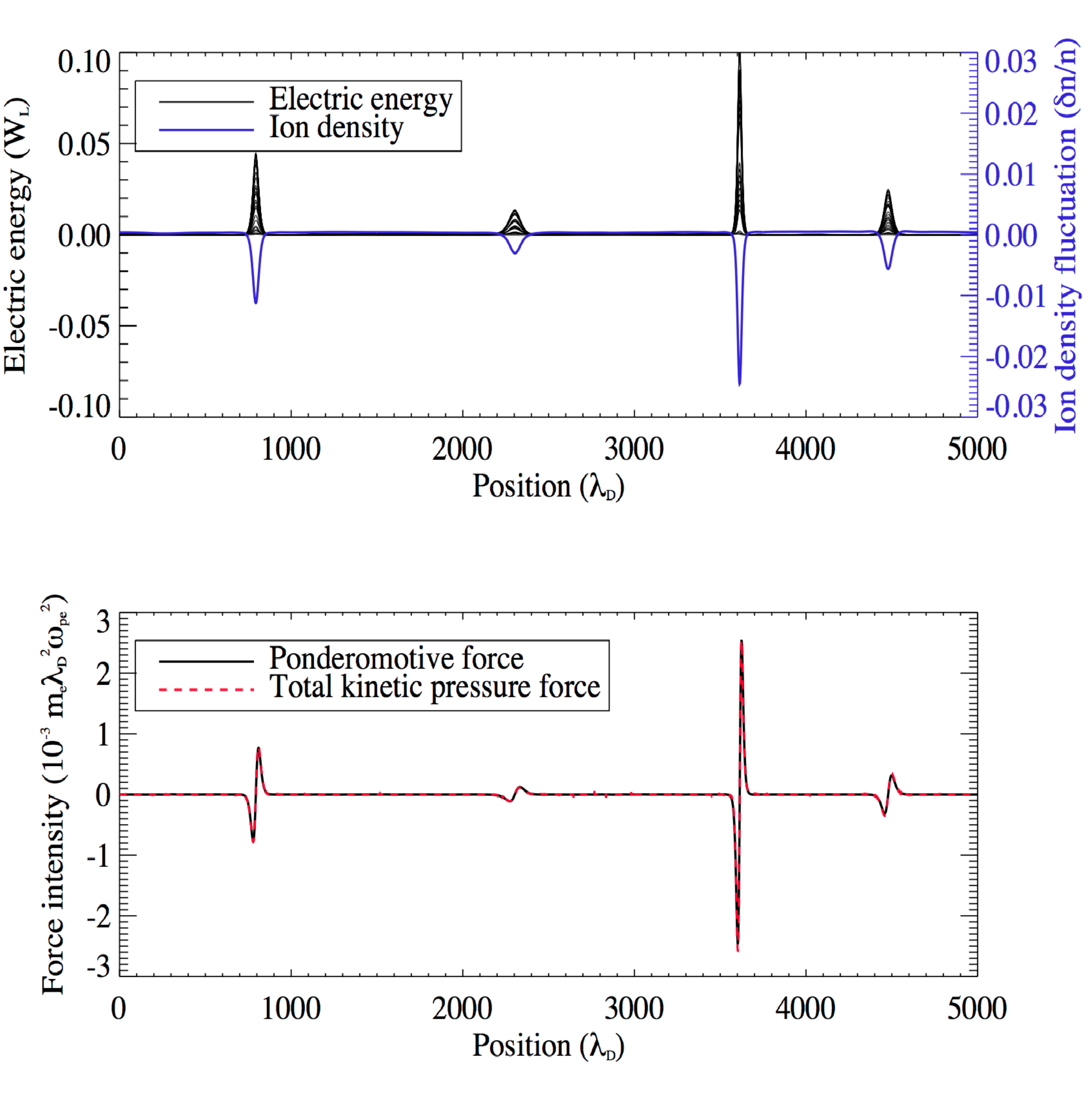}
\caption{Top panel: ion cavitons with the associated ion density fluctuations $\delta n / n$ (blue line) and electric energy $W_{_{L}}$ (black line). Bottom panel: The pressure and the ponderomotive force (red and black line, respectively). 
The ion density fluctuations as well as the envelop of the Langmuir electric energy remain practically constant from $t > 1.5 \times 10^5$.} 
\label{fig2}
\end{figure}

We first discuss the results of a simulation from the first set of numerical experiments, 
starting with a monochromatic L-wave with $E_{_{L}} = 0.06$ and $\lambda_{_{L}} = 100\ \lambda_{_{D}}$. 
The time evolution of the electric energy and the ion density are shown in Fig.~\ref{fig1}. The L-wave first undergoes electrostatic parametric instability (Langmuir electrostatic decay) during the period $3 \cdot 10^4  <  t < 6 \cdot 10^4$. In the mean time,  IA fluctuations are generated. 
Then, ion cavities start to form at $t \simeq 10^5$. These cavities are filled by  
electric energy in the form of an electrostatic field oscillating at the plasma frequency. 
The ion density fluctuations $\delta n / n$ and the Langmuir energy $W_{_{L}}$ are shown in Fig.~\ref{fig2}, top panel, (blue and black lines respectively) for $t > 1.5 \times 10^5$. The ion density fluctuations as well as the envelope of the Langmuir electric energy remain practically constant from $t > 1.5 \times 10^5$ to the end of the numerical simulation.
The cavities result from an equilibrium between the total pressure,  $-\nabla(P_{e} + P_{i})$ and the ponderomotive force, $- e^2/(4 m_{i} \omega_{pe}^2) \ \partial_{x} E^2$,  associated to high frequency Langmuir oscillations (see Fig.~\ref{fig2}, bottom panel, red and black line, respectively). These coherent structures, identified as "cavitons", are the signature of a transition to a strong Langmuir turbulent regime. \\

\begin{figure}[t!]
\noindent \includegraphics[width=\linewidth]{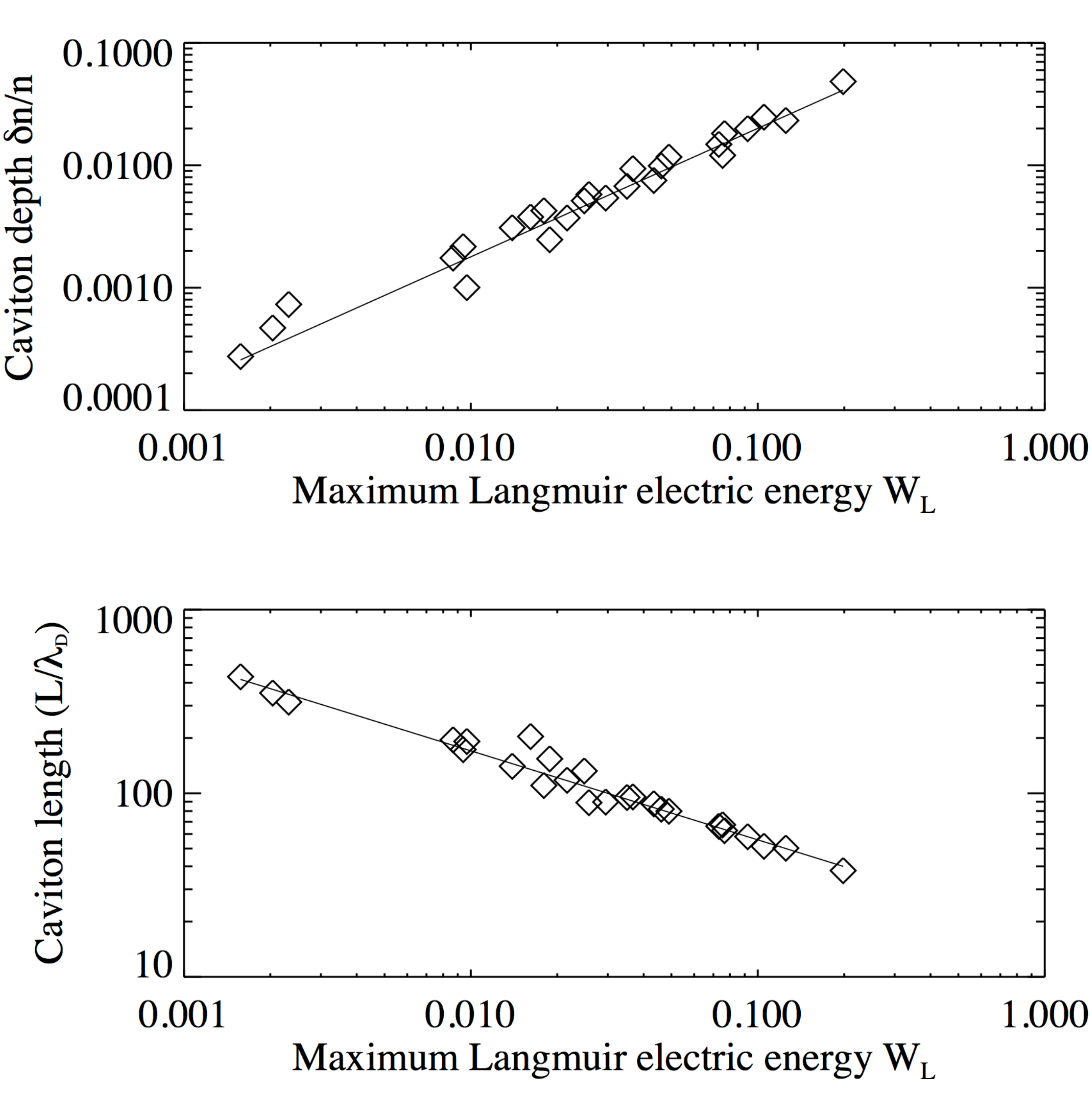}
\caption{Depth expressed in relative density $\delta n / n$ (top panel) and width $L$ expressed in Debye length (bottom panel) of cavitons measured in the simulations according to the associated Langmuir energy $W_{_{L}}$. Each diamond represents a single caviton. The line shows the power law fit.}
\label{fig3}
\end{figure}

We have performed many simulations starting with (i) coherent and (ii) incoherent initial L-waves. In all cases, the system freely evolves until it eventually jumps to a strong 
turbulence state characterized by the presence of Langmuir cavitons. The formation of 
cavitons does not need a high level of L-waves if the system evolves "long enough''. 
In case (i), as discussed, parametric decay occurs first and saturates in the first part of the simulation. Then cavitons are formed and remain stable until the end of the simulation.
In case (ii) the initial electron density fluctuation self-organises into a large spectrum of 
Langmuir noise. The "sea" of L-waves that fills the simulation box then collapses into stable low energy cavitons. 
In all cases the nonlinear structures are similar, indicating that the asymptotic behavior of 
Langmuir turbulence is independent from the initial level of coherence of Langmuir oscillations.
As expected, regardless of the initialization of the turbulence, the larger the initial L-waves, the sooner Langmuir cavitons are generated. 
For each caviton, we have measured the depth of the ion density hole $\delta n / n$, the length of the structure $L$ (width of the ion cavity at the height of $1/e$ of the maximum depth) and the  maximum Langmuir electric energy $W_{_{L}}$ that sustains the density hole. The results are shown in Fig.~\ref{fig3} on a  $W_{_{L}}$ interval ranging over three decades. It is worth noticing that Langmuir cavitons are observed also at remarkably low electric energy values, $W_{_{L}} \sim 10^{-3}$. The depth of the ion cavity $\delta n / n$ is of the order of the Langmuir electric energy density (as expected for high energy cavitons) over three decades:
\begin{equation} \label{eq:depthcavitons}
	\delta n / n = (\delta n / n)_{_{0}} \ W_{_{L}}^{\alpha}
\end{equation}
\noindent with $(\delta n / n)_{_{0}} = 0.28 \pm 0.06$ and $ \alpha = 1.13 \pm 0.06$ the fitting parameters. Fit results are given with their respective $3\sigma$ errors. 
The width of the cavitons also scales on the Langmuir electric density energy:
\begin{equation} \label{eq:lengthscalecavitons}
	L = L_{_{0}} \ W_{_{L}}^{\beta}
\end{equation}
\noindent with $L_{_{0}} = 18 \pm 4$ and $\beta = -0.47 \pm 0.05$. As expected, Langmuir cavitons have a larger scale length when the Langmuir electric energy is lower. 
A new unexpected and important result is the generation of stable Langmuir caviton with a scale length of many hundred of Debye lengths. 
Finally, the power law between the depth and the length of the ion cavities associated with the cavitons results as:
\begin{equation} \label{eq:lengthdepthcavitons}
L = L_{_{0}} \ (\delta n / n)^\gamma
\end{equation}
\noindent with $L_{_{0}} = 10 \pm 3$ and $\gamma = -0.42 \pm 0.05$. \\

 \section{Discussion} 

We hereafter discuss, first, the difference between the cavitons observed in our simulations and the Langmuir solitons also associated to strong Langmuir turbulence. Then we investigate the transition from weak to strong turbulence and its observed timescale. \\

\subsection{Langmuir solitons and cavitons}

Interestingly, the scaling laws obtained here for the Langmuir cavitons, Eq.~\ref{eq:depthcavitons}-\ref{eq:lengthdepthcavitons}, are similar to those of Langmuir solitons \cite{Nezlin1982}. In particular, the typical depth of the ion cavities is of the order of the Langmuir electric energy, obtained in the simulation over three orders of magnitude.
We recall that Langmuir cavitons, or Langmuir \emph{standing} solitons, are propagating structures maintaining their shape through the balance between dispersive and nonlinear effects, while cavitons are standing structures with electric oscillations that self-consistently sustain the density cavity of which they are eigenmodes.  \\

\begin{figure}[t!]
\includegraphics[width=\linewidth]{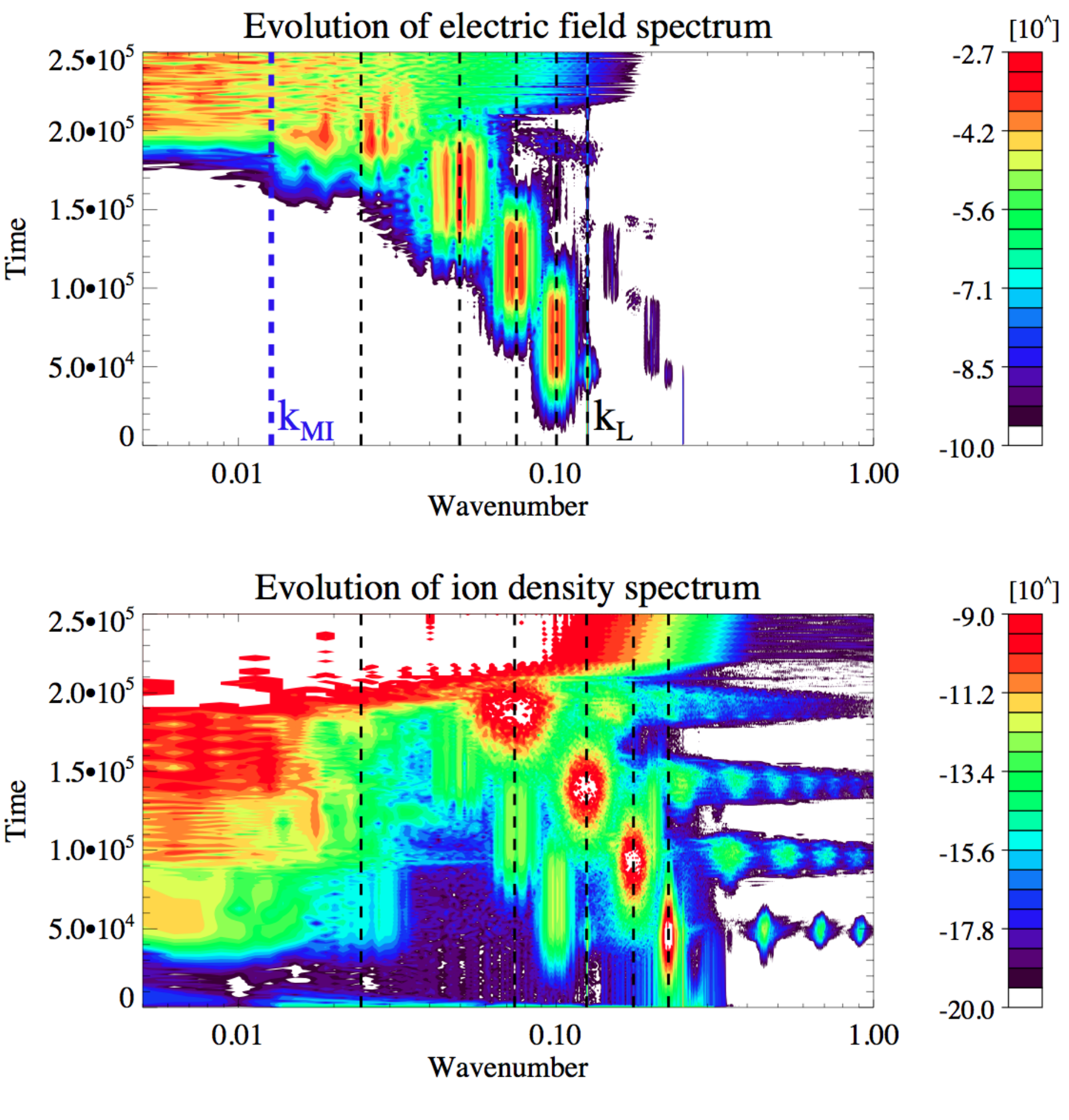}
\caption{\textbf{Evolution of the electric field and ion density spectrum (top and bottom panels respectively). Black dashed lines: expected wavenumber from 3-wave cascade interactions. Blue dashed line: transition wavenumber $k_{_{MI}}$ between decay and modulational  instabilities.}}
\label{fig4}
\end{figure}

\subsection{Transition from weak to strong turbulence}

In order to identify the mechanism responsible for the transition from weak to strong turbulence, we have studied the temporal evolution of the electric and ion density spectra. 
In the first part of the experiments, successive Langmuir decay instabilities generate smaller and smaller wave numbers (3-wave inverse cascade characteristic of weak turbulence). 

The 3-wave cascade spontaneously develops for the first series of numerical experiments starting from a monochromatic L-wave. This is shown in Fig.~\ref{fig4} (initial wave vector $k_{L} = 2 \pi / 50$) where we draw the development of the Langmuir turbulence (top panel) and the excitation of daughter S-waves (bottom panel) driven by the successive decays of the L-waves. The L-wave cascade follows the path in Fourier space expected for such 3-wave cascade process (the theoretical wave numbers are shown in dotted lines).
As previously pointed out \cite{Henri&al2010JGR}, the decay product S-waves are generated over a finite band of wave numbers and do not survive outside the coupling region with L-waves (bottom panel) because of strong Landau damping. Note that there is no cascade on the S-wave: instead the harmonics of the S-waves are temporarily excited as long as a pump L-wave injects energy into a fundamental S-wave, but the high kinetic damping prevents any further development of an inertial range on these S-waves.

Interestingly, the development of the L-wave cascade allows to identify isothermal electrons and adiabatic ions as the effective equation of state for ion acoustic fluctuations.
Here the ion sound speed $c_{s}$ is given by $c_{s}^2 = ( \gamma_{e} T_{e} + \gamma_{i} T_{i} ) / m_{i}$, where $\gamma_{e} =S 1$ and $\gamma_{i} = 5/3$. 
A slightly different definition of $c_{s}$ would lead to an incorrect path of the cascade in Fourier space. 
This effective equation of state, well known in the linear regime with $T_{e} >> T_{i}$, is thus extended to $T_{e} \sim T_{i}$ in the weak turbulence regime. 
This result is confirmed by a scattered plot of the ion pressure vs ion density during the weak turbulent stage of the simulation (not shown here). 

Once this weak turbulent cascade has developed, cavitons are systematically generated (apparition of a large spectrum in both panels) when $k_{_{L}} \lambda_{_{D}} \simeq 10^{-2}$ corresponding to a critical wave vector $k_{MI} \lambda_{_{D}} = 1/3 \ c_{s} / v_{th,e}$ (blue dashed line in Fig.~\ref{fig4}) for which the growth rate of the modulational instability, a known precursor of strong turbulence, overcomes the growth rate of the decay instability \cite{Zakharov&al1985PhR}. This is confirmed by other simulations starting with different initial wave vectors. 

To sum up, the weak turbulent inverse cascade brings the fluctuations to smaller wave vectors, where different nonlinear processes generate the coherent structures typical of a strong turbulent regime. In this picture, the strong turbulence regime is the natural asymptotic limit of weak turbulence. 

Finally, a series of experiments has been carried out with an initial wave vector $k_{L} < k_{MI}$, in order to directly trigger the modulational instability instead of the decay instability. No L-wave cascade is observed, but similar cavitons are indeed generated, in agreement with the previously exposed picture. 

The detailed mechanism for the generation of the cavitons is still unclear, even if it is known that cavitons can be by-products of the development of the modulational instability. This point will be investigated in future works. \\

\subsection{Evaluation of the nonlinear timescale}
\begin{figure}[t!]
\includegraphics[width=\linewidth]{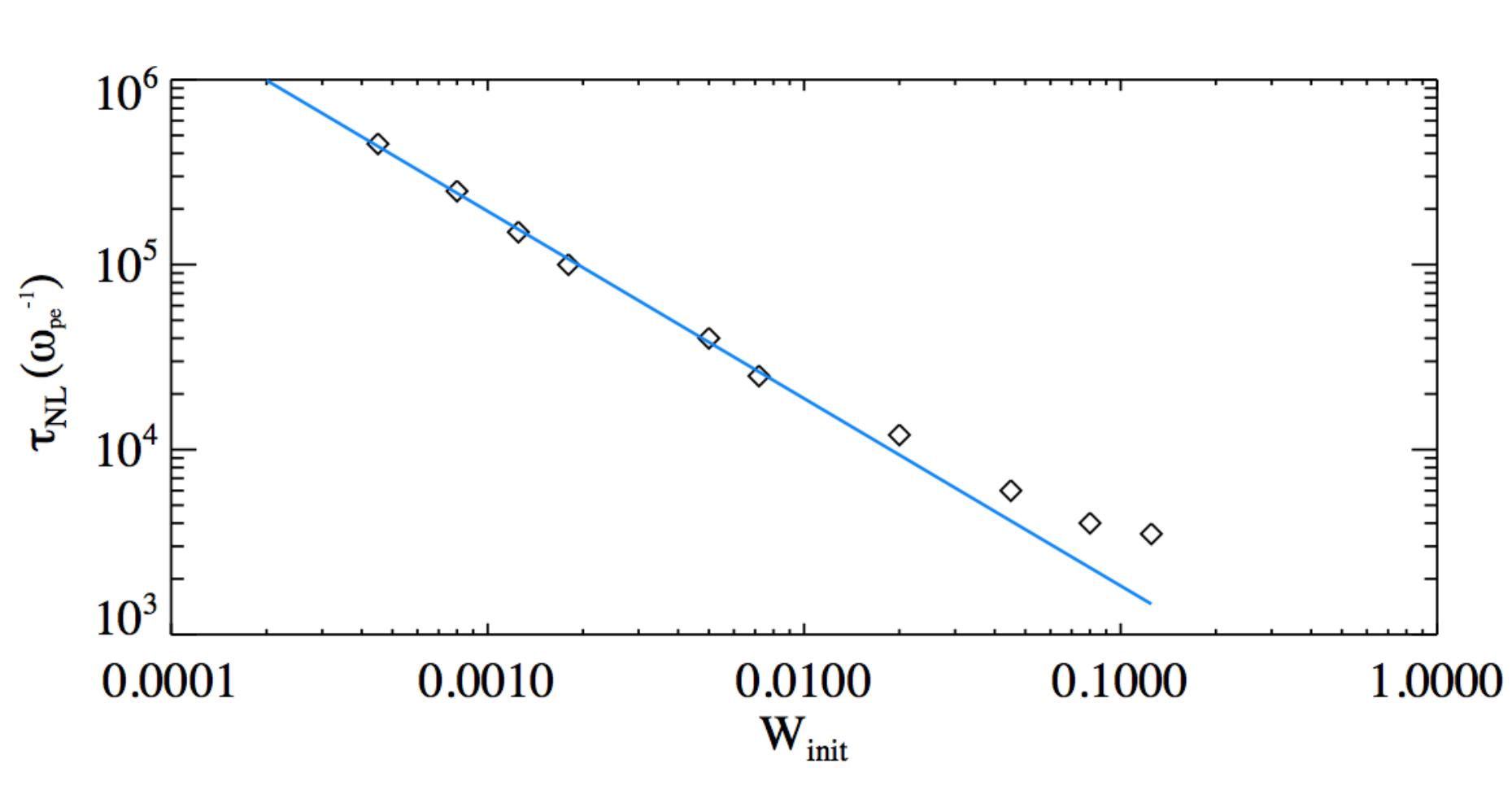}
\caption{The nonlinear time corresponding to the formation of Langmuir cavitons for different initial Langmuir energies $W_{_{L,init}}$. Diamond: simulation results, blue line: fit.}
\label{fig5}
\end{figure}

To complete this previous picture, we show in Fig.~\ref{fig5} the characteristic nonlinear time scale $\tau_{_{NL}}$ corresponding to the formation time of the first caviton in a simulation. We see that $\tau_{_{NL}}$ scales as the inverse of the initial Langmuir energy $W_{_{L,init}}$:
\begin{equation}
	\tau_{_{NL}} \simeq \tau_{_{NL} 0} \ W_{_{L,init}} ^{\eta} 
\end{equation}
\noindent with $\tau_{_{NL} 0} = 178 \pm 4$ and $\eta = -1.0 \pm 0.1$. 
Since the growth rate for Langmuir decay scales as $W_{_{L,init}}$ \cite{Bardwell&Goldman1976Apj}, $\tau_{_{NL}}$ is interpreted as the sum of a succession of Langmuir decay time scales to reach the critical wave vector $k_{c}$ for which the modulational instability becomes dominant and can generate the observed coherent structures (cavitons). 
Therefore, in a plasma like the solar wind, since in the L-waves are typically observed in the range $10^{-5} < W_{_{L}}^{obs} < 10^{-2}$, the time scale to reach a strong turbulence regime assuming $W_{_{L}}^{obs} \sim 10^{-2}$ would be about 10~sec (with $f_{pe} ~ 10$ kHz). \\

\section{Conclusions}

In summary, we have shown that L-wave turbulence is expected to breakdown for times larger than a typical nonlinear time scale, in agreement with Refs.~\cite{BivenNazarenkoNewell2001PhLA, BivenConnaughtonNewell2003PhyD}. 
The distinction between weak and strong turbulence thus looses part of its signification. These results are independent from the initial level of coherence. The formation of Langmuir cavitons appears to be the asymptotic limit of weak turbulence, as a result of an inverse Langmuir cascade.
The evolution of the electric spectrum shows that the weak turbulence regime brings itself the disturbances to large enough wavelengths for which the condensation of coherent structures becomes efficient, thus enabling the transition to strong turbulent features. 
The kinetic description allows to identify the effective equation of state for ion acoustic fluctuations in the weak turbulence regime of an electrostatic plasma with equal electron and proton temperatures. It is given by isothermal electrons and adiabatic ions; this extends the result known in the limit $T_e >> T_i$. 
Furthermore, electrostatic coherent structures of typical width much greater that a few Debye lengths are generated by the long time evolution of an initial relatively moderate amplitude turbulence. 
These results have been obtained for a 1D kinetic description of electrostatic plasmas and should be extended to multidimensional studies in order to test the stability of the observed coherent structures.

The breakdown of weak Langmuir turbulence and the existence of large coherent structures can have an important impact on the interpretation of space plasma data. The authors are thus confident that these new insights in Langmuir turbulence may encourage the space physics community to revisit the admitted conclusion that strong turbulent Langmuir structures are formed at too high energies to be relevant in space plasma environments. This last result should be directly tested on waveforms data in space plasma environments.  \\ \\ \\

\subsection*{Acknowledgments}
We are grateful to the italian super-computing center CINECA (Bologna) where part of the calculations where performed. We also acknowledge Dr.~C.~Cavazzoni for discussions on code performance.

\bibliographystyle{unsrt}
 \bibliography{/home/phenri/Documents/COMPTE_RENDUS/mabiblio}

\end{document}